\documentclass[aps,twocolumn,superscriptaddress,groupedaddress]{revtex4}  
\usepackage{graphicx}  
\usepackage{dcolumn}   
\usepackage{bm}        
\usepackage{amssymb}   
\usepackage{multirow}

\begin{document}

\title{Search for conformal invariance in compressible two-dimensional turbulence}

\author{S.~Stefanus} \affiliation{Department of Physics and Astronomy, University of Pittsburgh, Pittsburgh, Pennsylvania 15213, USA}
\author{J.~Larkin} \affiliation{Department of Mechanical Engineering, Carnegie Mellon University, Pittsburgh, Pennsylvania 15213, USA}
\author{W.I.~Goldburg} \affiliation{Department of Physics and Astronomy, University of Pittsburgh, Pittsburgh, Pennsylvania 15213, USA}

\vskip 0.25cm
\date{\today}

\begin{abstract}
We present a search for conformal invariance in vorticity isolines of two-dimensional compressible turbulence. The vorticity is measured by tracking the motion of particles that float at the surface of a turbulent tank of water. The three-dimensional turbulence in the tank has a Taylor microscale $Re_\lambda \simeq 160$. The conformal invariance theory being tested here is related to the behavior of equilibrium systems near a critical point. This theory is associated with the work of L\"owner, Schramm and others  and is usually referred to as Schramm-L\"owner Evolution (SLE). The system was exposed to several tests of SLE. The results of these tests suggest that zero-vorticity isolines exhibit noticeable departures from this type of conformal invariance.\\
Copyright (2011) American Institute of Physics. This article may be downloaded for personal use only. Any other use requires prior permission of the author and the American Institute of Physics. The following article appeared in (Stefanus, J. Larkin, W.I. Goldburg, "Search for conformal invariance in compressible two-dimensional turbulence," Phys. of Fluids {\bf 23}, 105101 (2011)) and may be found at http://link.aip.org/link/?PHF/23/105101.

\end{abstract}

\pacs{ go here}
\maketitle

\section{Introduction}

Strong turbulence is fiercely difficult to understand because it is dominated by nonlinear effects,  and because many degrees of freedom of fluid flow are excited \cite{landau}.  Some of its main features can be understood from dimensional arguments based on an eddy cascade theory.  There is a range of eddy sizes over which the system exhibits approximate self-similarity or scale invariance.  Most efforts have been focused on this  self-similar range, often called the inertial range.  It can span length scales ranging from many meters down to  eddy sizes of tens of microns in the atmosphere, in the ocean, and in large wind tunnels.    

Though most endeavors are focused on understanding three-dimensional turbulence, two dimensional (2D) flows are important from a practical and fundamental point of view.   The depth of oceans (L) and the thickness of the atmosphere is very small compared to the earth's radius $R$. Large-scale velocity variations $R>>L$ are properly viewed as two-dimensional. 

Two dimensional turbulence displays striking differences from its 3D counterpart. In 2D, smaller eddies  combine to form bigger ones while the reverse happens in three dimensions. This so-called "inverse cascade'' in 2D turbulence characterizes  hurricane growth in the troposphere \cite {Pedlosky}. 

The theory of two-dimensional (2D) turbulence brings new complications and simplifications at the same time. Between the large eddies in 2D turbulence are thin regions where the vorticity of the flow is very large, even though these regions contain only a small fraction of the turbulent energy. In two dimensions, the vorticity is, of course, perpendicular to the plane of the 2D flow and is hence a scalar.  In 3D, the vorticity is amplified  by velocity gradients, whereas in 2D, its mean square vorticity is a constant, viscous damping aside. As for this  damping, it occurs at small scales and is almost absent in the self-similar range of interest here.

It is obviously important if 2D turbulence should  turn out to exhibit invariance features that go beyond self-similarity.  Recent theoretical and numerical work suggests that this is so.   This evidence comes from the study  of contours of zero vorticity  in an incompressible flow \cite{Bernard} (to be called BBCF) .  In that important paper, the authors focus  on the geometry of (contorted) paths through the fluid where the vorticity $\omega$ is zero at each  instant of time $\tau$.  Their simulations reveal a new type of conformal invariance. The conformal invariance discussed here , and by BBCF, is unrelated to the usual conformal mapping technique applied to electric potential functions. Rather, it is about the growth of a random curve where each incremental length is produced by a conformal map characterized by a Brownian function.

A path in the $x,y$ plane, written as $z=(x,iy)$, is conformally invariant (in this sense \cite{Cardy}), if there is a function $g(z) =(u, iv) $ that can map the path back to the real axis in the $u,v$ plane while preserving all the angles (See Fig. \ref{sle_sample}).


The present work is an experimental study of  the contours of constant vorticity in a $compressible$ flow.  We know of no published experiments for incompressible systems. The present laboratory observations display approximate conformal invariance for the contours of zero vorticity measured at hundreds of instants of time $\tau$. Contours of nonzero vorticity paths were  also examined.   Larger deviations from conformal invariance are observed.

It was  noted by BBCF that with present technology, it is technically not possible to search for conformal invariance in $incompressible$ 2D flow experiments.  However, in the compressible flow experiments to be described here,  it was possible to accumulate data over a sufficiently wide parameter range to make such a test.  In this work, we use an overhead fast camera to track the motion of particles that float on a turbulent tank of water.  These particles have a density that is a fourth of the density of water, so their motion is confined to the surface of the underlying turbulent flow, which is, of course three-dimensional. The flow  of the  turbulent water underneath the floaters is incompressible, assuring that the two-dimensional divergence of the velocity of the floaters is not zero. The particles are small enough to be almost inertia-free.  Thus they sample the velocity of the flow ${\bf v}(x,y,z)$ in the surface plane $z=0$, as discussed below. The present experiments are performed at a moderately high Reynolds number where the inertial range is appreciably large.

This system of floaters is very different from conventional  two-dimensional turbulence and also from those studied by BBCF.  It is not merely that the surface on which the floaters move is rippled (their amplitude is small  \cite{Cressman}), it is that the floaters do not form a separate system; they can exchange energy with those water particles beneath them. In principle, at least, they can take and return their kinetic energy to the underlying fluid on all spatial scales.  Thus there is no reason to expect an energy cascade which implies dissipation only at small scales or at the boundaries of the container.  Nevertheless, the squared velocity difference $\langle \delta v({\bf r})^2 \rangle \equiv D_2(r)$ between pairs of points separated by distances $r$, closely conform to that of 3D turbulence, $D_2(r) \propto r^{2/3}$ .  This scaling is seen in experiments as well as simulations \cite {Cressman}.

The present study hinges on the measurement of a random variable $U(t)$, yet to be defined,  whose average mean square must be Brownian in character.  In that case
\begin{equation}\label{eq:Brown}
\langle (U(t) - U(0))^2 \rangle= \kappa t ^a .
\end{equation}
The exponent $a$ and the value of $\kappa$ are measured in the experiments discussed here.

The parameter $t$ is a dimensionless length and not time, and the exponent $a$ must be unity, as in Brownian motion, a requirement that must be met if conformal invariance  is realized.  The dimensionless "diffusivity" $\kappa$ is a very important parameter in the theory being tested here.  One may think of $\kappa$ as the dimensionless diffusivity, but only if $a$ = 1.   For a self-avoiding random walk $\kappa$ =8/3, and for critical percolation $\kappa$ = 6 \cite{Cardy}. This last value is deduced in the simulations of BBCF. In the present experiments,  the value of $\kappa$  for the zero-vorticity contours was extracted from measurements made at many instants of time.  At each instant of time $\tau$, there are many constant vorticity lines. The parameter $t$ increases along each line.

\section {The search for $U(t)$}

The analysis of the experimental data, to be discussed below, requires that the above-defined $g(z)$  be uniquely determined by the function $U(t)$ (usually referred to as the {\it driving function}), which is related to the experimental observations. Thus, $g(z)$ also depends on $t$ and hence will be written as $g_t(z)$.  Roughly speaking, the dimensionless parameter $t$ is proportional to the ``length'' of the 2D curve, which is being mapped by $g_t(z)$. This dependence on $t$ is what is usually referred to as the {\it L\"owner differential equation},
\begin{equation}\label{eq:Loewner}
\frac {\partial g_t(z)}{\partial t} = \frac {2} {g_t(z) - U(t)}.
\end{equation}

It was O. Schramm who first discovered that for a conformally invariant random curve in a 2D plane, $U(t)$ is  a one-dimensional Brownian motion obeying Eq. \ref{eq:Brown} \cite{Schramm}. In this case, the random curve is referred to as a Schramm-L\"owner Evolution (SLE) trace.  For a more formal and complete discussion of SLE, see \cite{Cardy}.
 
 One of the most effective ways to identify this type of conformal invariance is to measure $U(t)$, defined by the above differential equation, and see if its mean square average obeys Eq. \ref{eq:Brown}. In the next two sections, the experiment and the procedure to calculate $\langle (\delta U(t))^2 \rangle \equiv \langle (U(t) - U(0))^2 \rangle$ is described.
 
 \begin{figure}
\includegraphics[width=3 in]{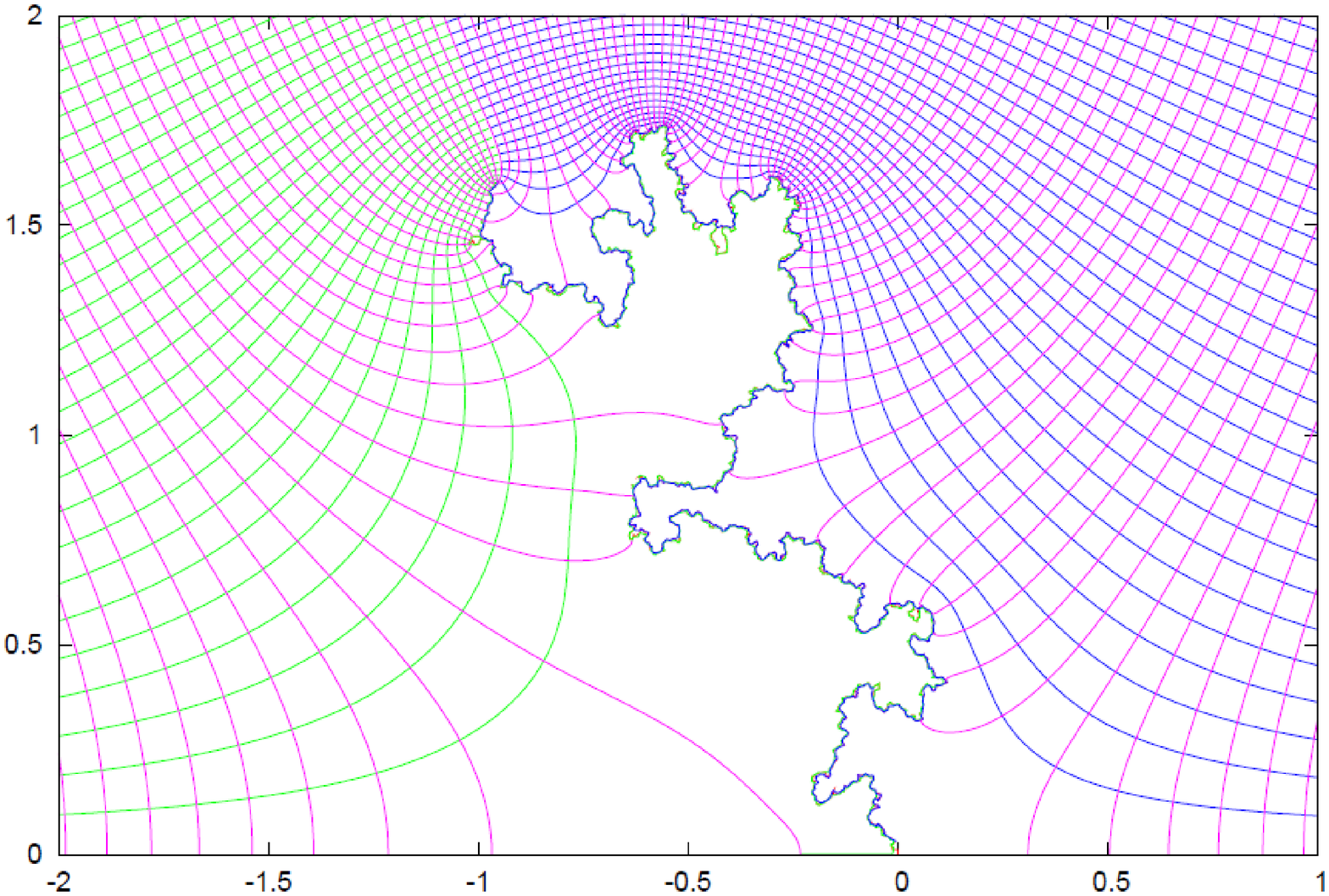}
\caption{A typical SLE trace with $\kappa=2$ is shown in Fig. 1. Writing $g_t(z)=u+iv, z=x+iy$,  the horizontal and vertical axes  in this figure are  $x$ and $iy$ respectively. The solid irregular line, which corresponds to a particular value the parameter $t$, separates a pair of regions where $g(z)$ is analytic. It is called a trace. The traces are self-similar and also self-avoiding. In the experiments to be discussed below, the traces were measured at  many of instants of (dimensionless) time $\tau$. (With permission from T. Kennedy. See T. Kennedy, http://www.math.arizona.edu/$\sim$tgk/rtg\textunderscore2011/sle2.0.pdf, 2011 for original graph.)}
\label{sle_sample}
\end{figure}

 A typical SLE trace is shown in Fig. \ref{sle_sample}.  Such traces are both self-similar and self-avoiding. After the conformal transformation, the grid in the $u,v$ plane is rectangular; the transformation under $g_t(z)$ is conformal, as all the angles are preserved.

\section {Experimental}
\begin{figure}
\includegraphics[width=3 in]{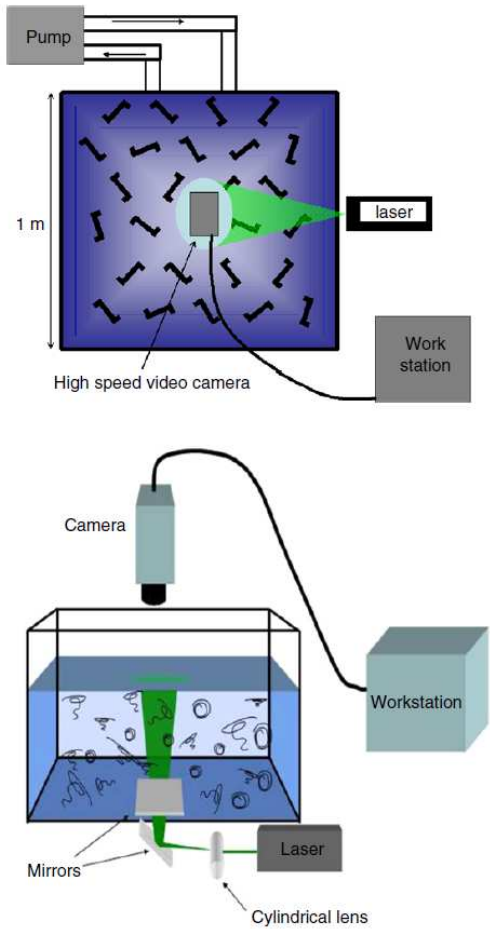}
\caption{Schematic of the top-view (top panel) and side-view (bottom panel) of the experimental setup. 36 rotating capped jets are placed horizontally on the tank floor (shown as randomly oriented Z-shaped patterns) that pump water into the tank recirculated by an 8 hp pump. The  region in the lateral center of the tank and at the surface ($z$=0) is illuminated by a laser-sheet. A high-speed digital camera suspended vertically above this central region captures images of the light scattered by buoyant particles (50 $\mu$m hollowglass spheres of specific gravity 0.25).}
\label{Setup}
\end{figure}

The 1m $\times$ 1m tank is filled with water to a height of 30 cm. The tank is large compared to the camera's field of view. The turbulence is generated by a large pump connected to a network of rotating jets in a plane 10 cm above the tank floor. See Fig. \ref {Setup} for a schematic of the experimental setup. The arrangement creates uniform turbulence in the center of the tank and also moves the source of turbulent injection far from the fluid surface where the measurements are made \cite{Cressman}. With this scheme, surface waves, which cannot be avoided, do not exceed an amplitude of $\sim$ 1 mm \cite {Cressman}. It is necessary that the surface of the tank be freshly cleaned before each set of measurements. Otherwise, amphiphiles form a continuous layer on the surface and prevents the floaters from moving freely under the action of the turbulence \cite {Cressman}. 

The hydrophilic particles chosen here are subject to capillary forces which are very small compared to forces coming from the turbulence, and do not affect the results as they do in \cite {Denissenko, Weinberg}. The non-inertial character of the particles is minimal because the Stokes number $St$ is small: $St \simeq 0.01$ \cite {Bandi}. 

During an experimental run, the floating particles (50 $\mu$m diameter and specific gravity of 0.25) are constantly seeded into the fluid from the tank floor, where they undergo turbulent mixing as they rise due to buoyancy and are uniformly dispersed by the time they rise to the surface. Once at the free-surface, their motion is constrained to the two-dimensional surface plane. Their motion is tracked with a high-speed camera (Phantom v.5) situated above the tank. The camera field-of-view is a square area of side length L = 9 cm. The constant particle injection is necessary to replace floaters that stick to the tank walls. The sources and sinks at the surface fluctuate in both time and space, which can cause particles to leave the camera's field of view. 

Instantaneous velocity fields are measured using an in-house developed particle imaging velocimetry (PIV) program which processes the recorded images of the floaters. The constant injection of particles ensures that surface sources and sinks receive an adequate coverage of particles on the surface. The local particle density at the surface determines the average spacing of the velocity vector fields produced by the PIV program. The resulting velocity vectors are spaced (on average) by $\delta x= 2.5 \eta$ over both sources and sinks , where $\eta$ is the size of the smallest eddies in the inertial range \cite{Larkin}.

\begin{table}
\caption{\label{param}Turbulent parameters measured at the surface. Measurements are made at several values of $Re_\lambda$ with an average $Re_\lambda \simeq 160$. The parameters listed are averages, with deviations less than 10\%.}
\begin{ruledtabular}
\begin{tabular}{l | c | c }
\multirow{2}{*} {}Parameter & Symbol used & Measured \\
& in text & value \\
\hline  & & \\
Taylor microscale $\lambda$ (cm) & $\lambda = \sqrt{\frac{v_{rms}^2}{\langle (\partial v_x / \partial x)^2 \rangle}}$ & 0.37 \\ & & \\
Taylor $Re_\lambda$ & $Re_\lambda = \frac{v_{rms}\lambda}{\nu}$ & 160\\  & & \\
Integral scale $l_0$ (cm) & $l_0 = \int dr \frac{\langle v_{\Vert}(x+r)v_{\Vert}(x)\rangle}{\langle (v_{\Vert}(x))^2 \rangle}$ & 1.42 \\ & &  \\
\multirow{2}{*} {}Large Eddy Turnover & $\tau_0 = \frac{l_0}{v_{rms}}$  & 0.43 \\
Time (LETT) $\tau_0$ (s)& & \\ & &  \\
\multirow{2}{*} {}Dissipation rate &  $\varepsilon_{diss}=10\nu \langle (\frac{\partial v_x}{\partial x})^2 \rangle$ & 6.05 \\
$\varepsilon_{diss}$ (cm$^2$/s$^3$)  & & \\ & &  \\
Kolmogorov scale $\eta$ (cm) & $\eta = (\frac{\nu^3}{\varepsilon})^{1/4}$ & 0.02 \\ & &  \\
RMS velocity $v_{rms}$(cm/s) & $v_{rms}=\sqrt{\langle v^2 \rangle- \langle v \rangle^2}$ & 3.3 \\  & & \\
Compressibility $\mathcal{C}$ & $\mathcal{C} = \frac{\langle (\vec{\nabla_2}\cdot \vec{v})^2 \rangle}{\langle (\vec{\nabla_2} \vec{v})^2 \rangle}$ & 0.49 $\pm$ 2\%\\

\end{tabular}
\end{ruledtabular}
\end{table}

Data were taken for several values of $Re_\lambda \simeq 150-170$ with an average $Re_\lambda \simeq 160$.  Turbulent parameters measured at the surface are listed in Table \ref{param}. All of the statistics presented below were obtained by evolving $\sim 10^5$ Lagrangian particles in each frame.  
\section {Results and Discussion}

\begin{figure}
\includegraphics[width=3 in]{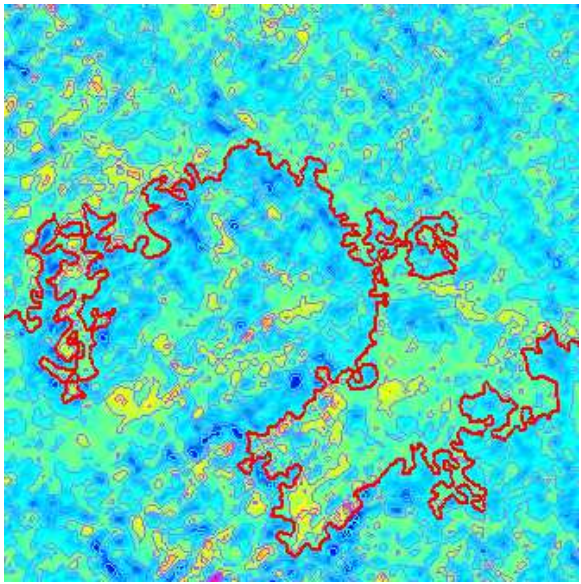}
\caption{A typical vorticity field displayed with the isolines. The image is of a square 9 cm by 9 cm. The thick solid line is the longest zero isoline in this particular field.}
\label{sample}
\end{figure}

\begin{figure}
\includegraphics[width=3 in]{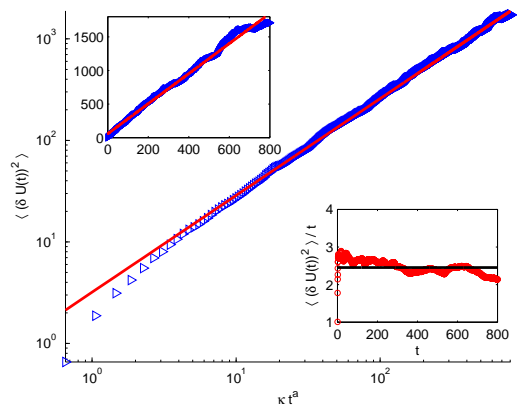}
\caption{Main frame: a typical log-log plot of $\langle (U(t + t_0) - U(t_0))^2 \rangle$ showing its linearity with respect to $t$. Upper-left inset: a typical linear plot of $\langle (U(t + t_0) - U(t_0))^2 \rangle$ showing the value of $\kappa$ as its slope. Lower-right inset: the distribution of $\kappa = \langle (\delta U(t))^2 \rangle/t$}
\label{Fitting}
\end{figure}

The vorticity field at each point, and at each instant of time $\tau$, is extracted from a measurement of the velocity field at equally spaced points separated by $\simeq 2.7 \eta = 0.54$ mm over the 9 cm $\times$ 9 cm field of view of the camera, giving a resolution of 160 $\times$ 160 points. The camera captures the images at a  rate of 133 Hz. The total number of frames that were analyzed is 833.

In each frame there are roughly 1000 vorticity isolines and 400 zero-vorticity lines, see Fig. \ref{sample}. The vorticity field is approximated using Stoke's Theorem $\omega = \frac {\oint {\bf v} \cdot {\bf dl}} {A}$, where $A$ is an area of size $\simeq (2.7 \eta)^2$. Although for a fractal object the concept of length is not clearly specified, the average length of the longest zero isoline of each frame is roughly 1000 steps, where the step size is $1.35 \eta$.

From each frame, the longest zero and non-zero vorticity lines are extracted. An x-y axis is chosen for each isoline in each frame. The lowest point of the isoline is taken as the origin of the complex plane, $x = y = 0$  because  $g_t(z)$ only concerns the upper complex plane, $y > 0$. Next, we calculate the driving function $U(t)$ of each of the vorticity isolines using an algorithm implemented by T. Kennedy \cite {Kennedy}.

Stated briefly, the algorithm goes by decomposing $g_{s+t} (z)=\bar{g}_t \circ g_s (z)$, where $g_s(z)$ maps the curve from 0 to $s$ back to the real axis and $\bar{g}_t (z)$ maps the image of the curve from $s$ to $t$ under $g_s(z)$ back to the real axis. By recursively decomposing $g_s(z)$, we get $g_{t_k} = \bar{g}_k \circ \bar{g}_{k-1} \circ \bar{g}_{k-2} \circ ... \circ \bar{g}_2 \circ \bar{g}_1, 0 = t_0 < t_1 < t_2 ... <t_k$. Each $\bar{g}_k$ maps one small segment $\Delta t$ of the curve to the real axis, where $\bar {g}_k$ has to satisfy L\"owner equation (Eq. \ref{eq:Loewner}) within each $\Delta t$. In principle, we take $\Delta t \rightarrow 0$ but in the experiments the smallest $\Delta t$ is the spatial interval of the two closest points with the same vorticity ($\Delta t \simeq 2.7 \eta$). Each incremental map $\bar{g}_k(z)$ is approximated by a Laurent series around $z \rightarrow \infty$, $\bar{g}_k(z) = z + \frac{\Delta t}{z}  + U_k + O(\frac{1}{z^2})$. This produces a sequence of discrete $U_k$ that approximates the true driving functions $U(t)$.

There are two ways of calculating $\langle (\delta U(t))^2 \rangle$ (Eq. \ref{eq:Brown}). One way is to do self-averaging along $t$ followed by an ensemble average over the 833 runs, another way is to calculate the ensemble average over $t$ directly as done by BBCF. If $U(t)$ is Brownian, as expected of SLE traces, the two methods should yield the same result for $\kappa$ and the exponent $a$. The resulting values of $\kappa$ and $a$ from the two different methods will be denoted $\kappa_t$ and $a_t$ for the self-average and $\kappa_e$ and $a_e$ for the ensemble average.

The self-averaging is done by calculating $\langle (U(t + t_0) - U(t_0))^2 \rangle$, where $\langle \cdot\cdot\cdot \rangle$ is an average over $t_0$ for each fixed value of $t$. The exponent $a$ in Eq. \ref{eq:Brown}  is calculated by taking the log of both sides of the equation, i.e. $\log \langle (U(t + t_0) - U(t_0))^2 \rangle = a \log (t) + \log (\kappa)$ for each curve and fitting it to a straight line (Fig. \ref{Fitting}). The slope of the line is the value of the exponent $a$ for that particular curve. The ensemble average over the 833 values of $\kappa$ and $a$ are then taken to be the value for $\kappa_t$ and $a_t$ for the system.

For SLE, as for Brownian motion, these two methods of averaging should produce the same results. Therefore, an easy way to distinguish vorticity data from effects of random noise follows from  comparing results obtained by these two different averaging methods.  To illustrate, we simulated SLE traces, and performed this particular test on those traces for a range of $\kappa$-values from 1 to 8 with increments of 0.5. In this simulation, there are 1000 SLE traces for each value of $\kappa$; each trace  contains  4000 points. We compare the results of $\langle (\delta U(t))^2 \rangle$ of the two averaging methods and find that they yield the same values of $\kappa$ within  2 \%.

We also extracted isolines from purely random fields. These fields were produced by generating uniformly distributed vorticity amplitude at each site and have the same resolution as the measured velocity fields. The isolines of the random fields satisfy other tests of SLE (as described below) but fail the ensemble vs self average comparison test. Thus, comparing the results of the two averaging methods proves to be a very useful tool for differentiating noise effects from real SLE traces.

Turning now to our experimental data, as seen in Table \ref{tab:comparison}, the values produced by the two averaging procedures differ by approximately two standard deviations for both $\kappa$ and $a$. Since $a_t$ and $a_e$ are measurably different from unity, the meaning of the $\kappa$'s in this case is ambiguous.   We defer discussion of other parameters in the table.   

\begin{figure}
\includegraphics[width=3 in]{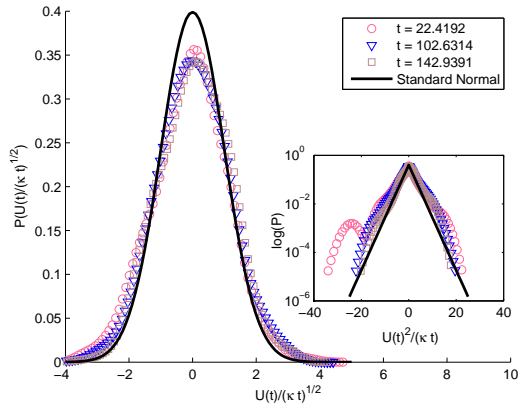}
\caption{Main frame: the probability distribution function of $U(t)/(\kappa t)^{1/2}$ for the zero-vorticity isolines for three different values of t. Lower-right inset: the log-linear plot of the PDF for 3 different values of $t$.}
\label{Driving}
\end{figure}

\begin{table}
\caption{\label{tab:comparison}Results of different tests of conformal invariance for zero isolines.  For the non-zero isolines the results show that they are not conformally invariant}
\begin{ruledtabular}
\begin{tabular}{l|c|c|c}
\multirow{2}{*} {} Test & $t$ Avg & Ensemble Avg & $SLE_\kappa$  \\
 & (833 runs) & (833 runs) \\
\hline & & & \\
Exponent $a$ & $a_t$=1.2 $\pm$ 0.1 & $a_e$=0.97 & $a_t$=$a_e$=1 \\  \hline & & & \\
$\kappa$ & $\kappa_t=$3.9 $\pm$ 1 & $\kappa_e=$2.45 $\pm$ 0.2 & $\kappa_t=\kappa_e$ \\ \hline
\multirow{3}{*} {}&  & Not gaussian & \\
$P(U(t)/\sqrt{\kappa t})$  & & see Fig.\ref{Driving} & $\propto e^{{-x^2}/2}$ \\ 
& & and text & $\langle x \rangle = 0$ \\
\hline
\multirow{3}{*} {}$P(\theta)$ Eq. \ref{eq:Hyper} & & & \\
see dotted line &  No & Yes & see text\\ 
in Fig. \ref{Pdf} & & & \\
\hline 
\multirow{3}{*} {}$D_q$ independent & & & \\
 of $q$, $q$=0 to 10 & see text & No & Yes\\
see Fig. \ref{Dq} & & & \\ 

\end{tabular}
\end{ruledtabular}
\end{table}

Two stringent tests of SLE, namely $\chi_a^2$ and $\chi_b^2$ as described by Kennedy \cite{Kennedy08}, were also performed on the isolines.  The experimental data failed this test, and so did a simulated SLE at a similar resolution.  Clearly a more refined resolution is required to produce a conclusive result.

The function $U(t)$ is a random variable, and we have measured its probability distribution function (PDF, see Fig. \ref{Driving})  to determine if it is gaussian for all values of $t$ as required for a Brownian process. If so, the PDF of $U(t)/(\kappa_e t)^{1/2}$ should collapse onto a standard gaussian PDF with mean $\mu = 0$ and standard deviation $\sigma = 1$ (Fig. \ref{Driving}). Here, only $\kappa_e$ is meaningful, since the PDF of $U(t)$ is produced by the values of $U(t)$ for a fixed $t$ in the ensemble. The inset of Fig. \ref{Driving} shows that the PDF's of $U(t)$ for three different values of $t$ do not conform to a gaussian distribution. That is, all data points do not lie on an inverted V. For the non-zero isolines the PDF of $U(t)/(\kappa_e t)^{1/2}$ has non-zero mean and is strongly skewed, ruling out conformal invariance (mean and skewness of the zero isolines are 0.04 and -0.0006 respectively while for the non-zero isolines they are 0.6 and -0.2 respectively).


\begin{figure*}
\includegraphics[width=3 in]{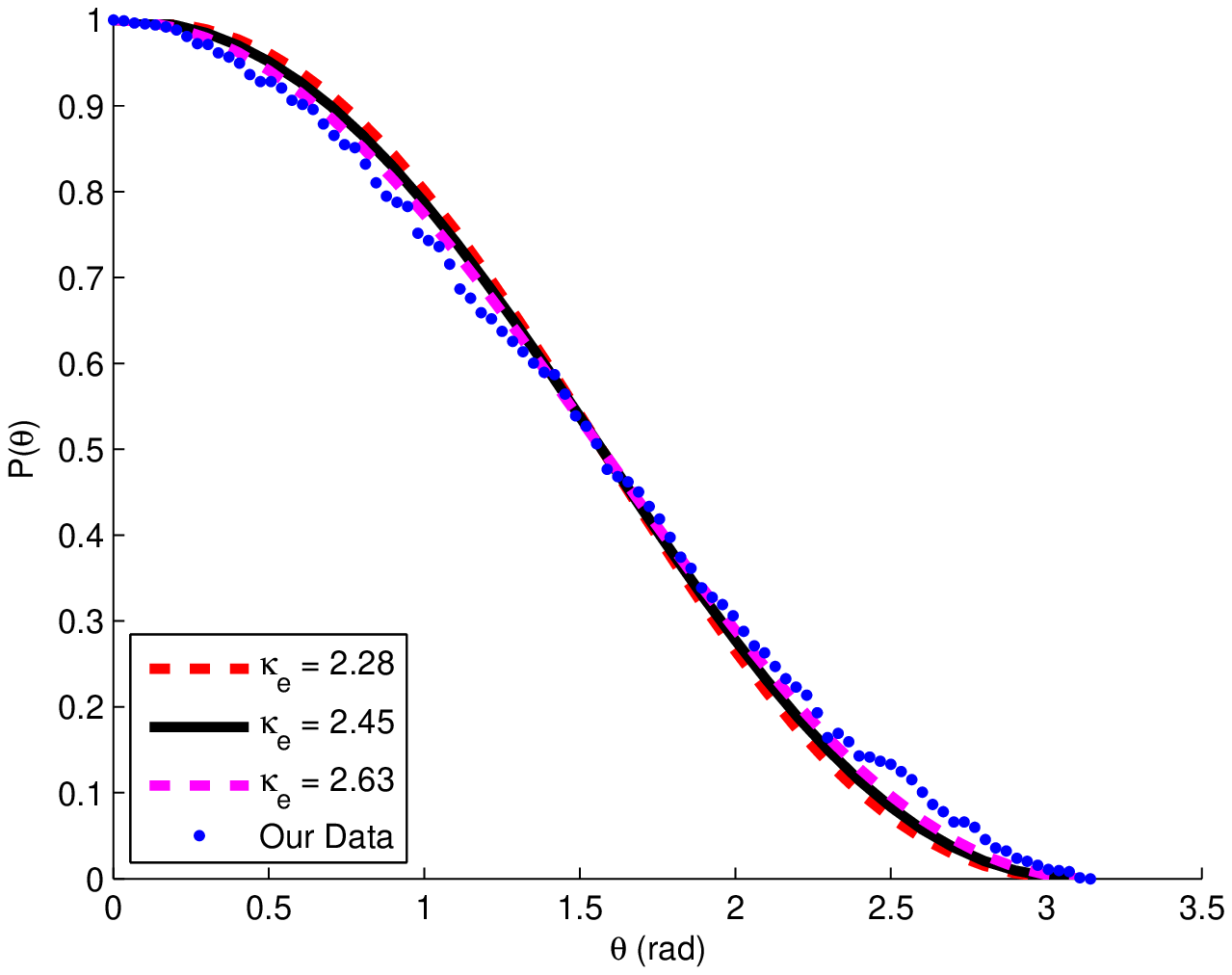}
\includegraphics[width=3 in]{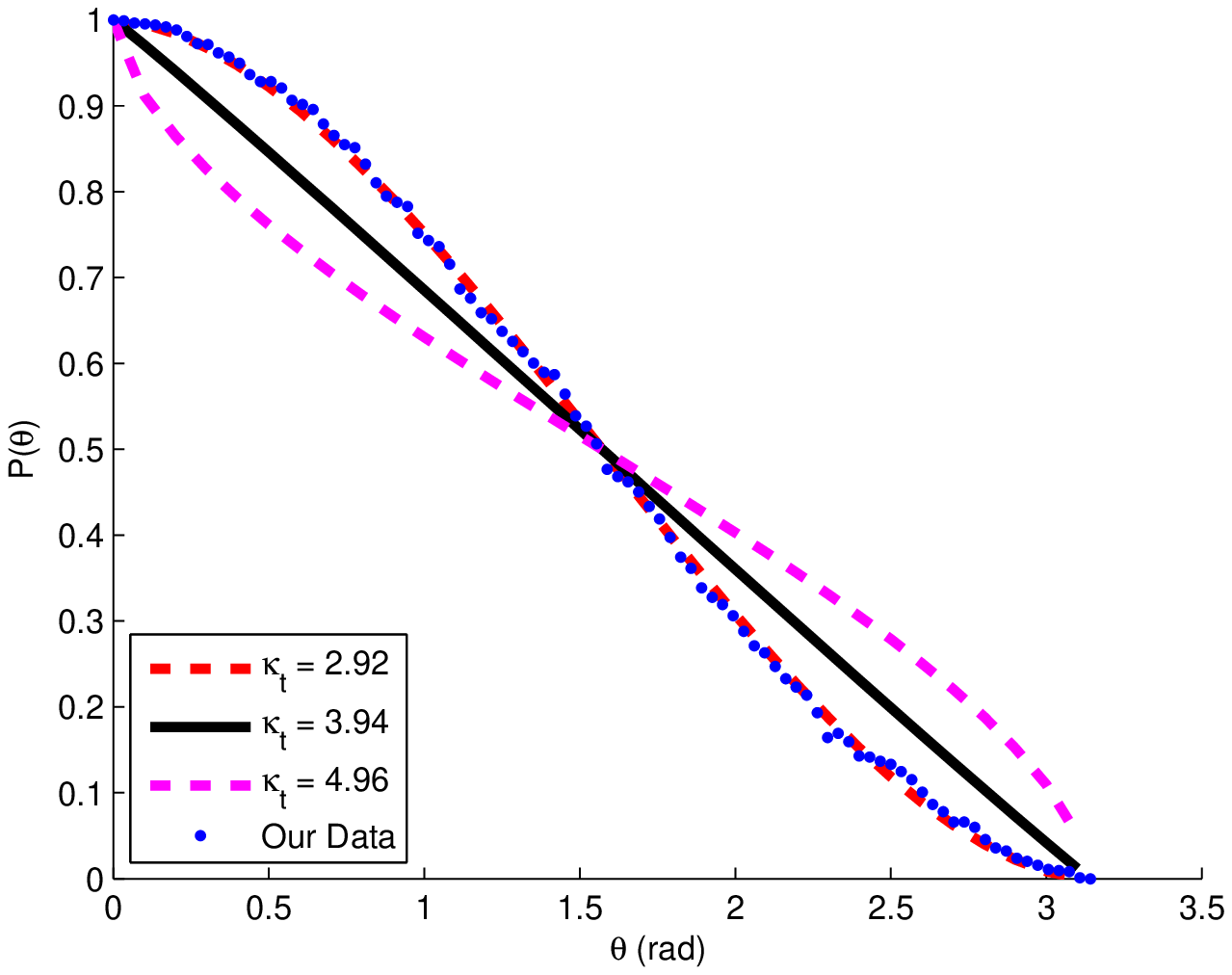}
\caption{Probability that an isoline keeps a given $z = \rho e^{i \theta}$ to its right. The horizontal axis is $\theta$ in radians with respect to the x axis. The probability distribution for zero  isolines compared to the expected values for $\kappa_e$ (Left) and $\kappa_t$ (Right). Solid lines are predicted distribution based on mean of $\kappa$ and dashed lines are based on the value of $\kappa \pm$ one standard deviation.}
\label{Pdf}
\end{figure*}

We apply another test to validate the Brownian characteristic of $U(t)$ (Eq. \ref{eq:Brown}). If the vorticity isolines are characterized by Brownian $U(t)$, they must be identified by a particular probability distribution for keeping a point $z = \rho e^{i \theta}$ to the right of each isoline. Here $\theta$ is defined with respect to the x axis. It is given by \cite {Kager}
\begin{equation}\label{eq:Hyper}
P(\theta) = \frac {1}{2} + \left ( \frac {\Gamma \left ( \frac {4}{\kappa} \right )} {\sqrt{\pi} \Gamma \left ( \frac {8 - \kappa} {2 \kappa}\right )} \right ) {}_2F_1 \left ( \frac{1}{2},\frac {4}{\kappa};\frac {3}{2};-\cot^2\theta \right ) \cot (\theta),
\end{equation}
where $\Gamma$ is the ordinary Gamma function with $\kappa$ as a parameter, and ${}_2F_1$ is the gauss hypergeometric function.

The probability distribution for the zero isolines is shown in Fig. \ref{Pdf}. The solid lines represent the distributions based on the mean values of $\kappa$; the dashed lines denote $\pm$ one standard deviation. The dots are the measured distribution. The left panel of this figure shows better agreement with our measurements. Equivalently, the measured angular distribution fits better to the expected distribution for $\kappa_e$. We are puzzled by this finding.



\begin{figure}
\includegraphics[width=3 in]{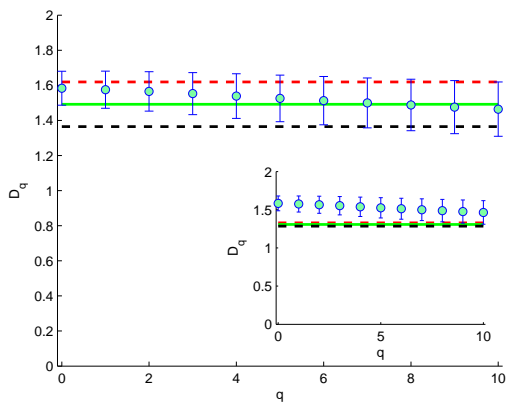}
\caption{Main frame: multifractal spectrum of the zero vorticity isolines compared to the value of $D_q$ based on $\kappa_t$. Lower-right inset: multifractal spectrum of the zero vorticity isolines compared to the value of $D_q$ based on $\kappa_e$}
\label{Dq}
\end{figure}

Our last test to see whether the system exhibits conformal invariance is calculating the multifractal spectrum of the isolines (Fig. \ref{Dq}), since conformal invariance requires scale invariance. The multifractal spectrum $D_q$ of the longest zero and non-zero vorticity isoline of each frame is then computed using \cite {Hilborn}
\begin{equation}
D_q = \lim_{q \to 0} \frac {1} {q-1} \frac {d \log (C_q(r))} {d \log r}
\end{equation}
\begin{equation}\label{eq:Cr}
C_q(r) = \frac {1}{N} \sum_{i}^{N}  \left [ \frac {1} {N-1} \sum_{j \not= i}^{N-1} H (r - r_{ij}) \right ]^{q-1}
\end{equation}
Here N are the total number of points in the isoline, $H$ is the heaviside step function, and $r_{ij}$ is the distance between points $i$ and $j$. This algorithm for determining the spectrum of fractal dimensions is given by Hentschel and Procaccia \cite {Hentschel}.

To calculate $D_q$, the log of the correlation sum (Eq. \ref{eq:Cr}) is plotted versus the log of r. The range of r over which the plot is a straight line is the scale-free (or scaling) region. The slope of the line $d \log (C_q(r))/d \log(r)$ is the value of $D_q$.

\begin{figure}
\includegraphics[width=3 in]{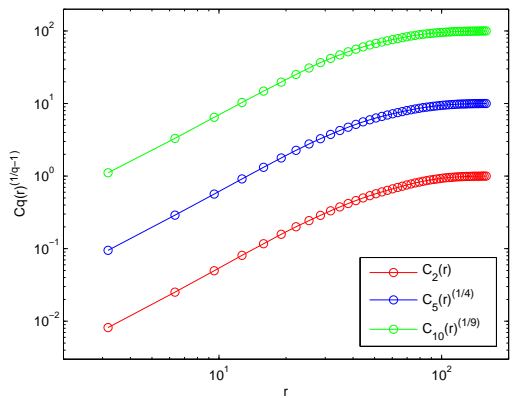}
\caption{Log-log plot of typical $C_q(r)^{1/q-1}$ versus $r$ for $q$=2, 5, and 10.  The scaling region (straight portion) of $C_q(r)$ typically extends 0.7 decades in $r$ as depicted in the plot. From top to bottom, the curves are $C_{10}$, $C_{5}$, and $C_{2}$}
\label{CqR}
\end{figure}

There are three lines in each graph in Fig. \ref{Dq}. They are the fractal dimensions given by $D_\kappa = 1 + \frac {\kappa}{8}, \kappa < 8$ \cite {Saleur,Beffara}. The middle horizontal line in each graph represents $D_\kappa$ using the measured mean value of $\kappa$ (taking $a$ to be unity); the upper and lower dashed horizontal lines show $D_\kappa$ using one standard deviation from the mean value of $\kappa$. The dots represent the measured mean fractal dimension for each $q$. The vertical error bars show the uncertainty in the measured values of $D_q$. The scaling region of $C_q(r)$, of which slope determines the value of $D_q$, is shown in Fig. \ref{CqR}. The scaling region typically extends 7/10 of a decade in $r$ for both $q$ = 2 and $q$ = 10.

The multifractal spectrum of the zero isolines is then compared to that of the expected value of $D_\kappa$. Fig. \ref{Dq} shows that the multifractal spectrum of the zero isolines conforms better to the value of $D_\kappa$ given by $\kappa_t$. It is interesting to note that $P(\theta)$ fits better to the distribution given by $\kappa_e$. Strictly speaking, a conformally invariant curve should have a constant multifractal spectrum which is independent of $q$.

In light of the error bars in this figure, we are forced to conclude that departures from homogeneous fractal behavior are not clearly present.  At the same time, the secular decrease in  $D_q$ with increasing $q$, suggest that $D_q$ may not be a homogeneous fractal. 



%

\section {Summary}

The goal of this experiment is to determine if the vorticity isolines of the compressible system of floaters meet all the imposed tests for SLE.  The floaters exhibit a measurable departure  from SLE for the zero vorticity isolines and a much larger departure for lines of nonzero vorticity. The departures are clearly evident in  Fig. \ref{Driving}. The interest of this study may be that the floaters display approximate SLE, as Fig. \ref{Fitting} and Table \ref{tab:comparison} show.



\section {Acknowledgements}

We would like to thank Tony Roberts for his laborious effort in re-calculating the $D_q$ for one of our data sets and Tom Kennedy for his SLE code and Fig. \ref{sle_sample}.

\end{document}